\documentclass[acmtog]{acmart}

\AtBeginDocument{%
  \providecommand\BibTeX{{%
    \normalfont B\kern-0.5em{\scshape i\kern-0.25em b}\kern-0.8em\TeX}}}

\begin{document}



\title{Towards a New Science of Disinformation}

\author{Claudio S. Pinhanez}
\authornote{Authors listed in alphabetical order}
\authornote{Principal Investigator}
\email{csantosp@br.ibm.com}
\affiliation{%
  \institution{IBM Research - Brazil}
}

\author{German H. Flores}
\authornote{Work done at IBM Research}
\affiliation{%
  \institution{IBM Research - Almaden}
}
\email{ghflores@us.ibm.com}

\author{Marisa A. Vasconcelos}
\affiliation{%
 \institution{IBM Research - Brazil}
}
\email{marisaav@br.ibm.com}

\author{Mu Qiao}
\authornotemark[2]
\affiliation{%
  \institution{IBM Research - Almaden}
}
\email{mqiao@us.ibm.com}

\author{Nick Linck}
\authornotemark[3]
\affiliation{%
  \institution{IBM Research - Almaden}
}
\email{Nick.Linck@ibm.com}

\author{Rog\'erio de Paula}
\authornotemark[2]
\affiliation{\institution{IBM Research - Brazil}}
\email{ropaula@br.ibm.com}

\author{Yuya J. Ong}
\affiliation{\institution{IBM Research - Almaden}}
\email{yuyajong@ibm.com}


\begin{abstract}
{\bf Abstract:} How can we best address the dangerous impact that deep learning-generated fake audios, photographs, and videos (a.k.a. \emph{deepfakes}) may have in personal and societal life? We foresee that the availability of cheap deepfake technology will create a second wave of disinformation where people will receive specific, personalized disinformation through different channels, making the current approaches to fight disinformation obsolete. We argue that \emph{fake media} has to be seen as an upcoming cybersecurity problem, and we have to shift from combating its spread to a prevention and cure framework where users have available ways to verify, challenge, and argue against the veracity of each piece of media they are exposed to. To create the technologies behind this framework, we propose that a new \emph{Science of Disinformation} is needed, one which creates a theoretical framework both for the processes of \emph{communication} and \emph{consumption} of false content. Key scientific and technological challenges facing this research agenda are listed and discussed in the light of state-of-art technologies for fake media generation and detection, argument finding and construction, and how to effectively engage users in the prevention and cure processes.
\end{abstract}

%


\maketitle

\section{Introduction}

In the past 20 years, we have witnessed the emergence and widespread dissemination of fake news in social media, which have been extensively reported and whose impact magnitude has been enormous, notably in the political, health, and celebrity domains. Since the 2016 US Presidential election and more recently with the COVID-19 Pandemic, when the pervasiveness and influence of fake-news achieved a scale and reach not experienced before, the investigations and studies of misinformation and disinformation became part of a prominent research agenda in computer science, social sciences, journalism, and political science (e.g., \cite{grossman2019electoral, allcott2017social, Lazer1094, Bessi_Ferrara_2016, chesney2019deepfakes, Vosoughi1146, wang2019systematic, pennycook2020fighting} to list just a few). We nevertheless characterize it as the \emph{first wave of disinformation}. With the recent addition of \emph{deepfake} methods and tools, which employ deep-learning techniques, particularly generative adversarial networks (GAN), for fake material generation, we are gearing up for the \emph{second wave of disinformation} based on extremely realistic deceptive depictions of people and events in text, audio, photos, and video, which we refer here as \emph{fake media}. 

The erosion of trust can be thought of as the main perverse impact of \emph{deepfakes} on people, organizations, and society at large. Trust lies at the heart of any human interaction, being the cornerstone of any social contract and business operation. It is the oil that lubricates the economy engines. The fact that \emph{deepfakes} challenge our capacity of  distinguishing between true and false information makes it extremely dangerous to the society as it blurs our perception of true reality. The mere existence of such a capacity is enough to transform fundamentally the ways in which we carry our everyday affairs, do business, and participate in society (see \cite{chesney2019deepfakes}). The  existence of deepfakes is enough to bring more credibility to denials -- \emph{the Liar's dividend}. The spread of harmful information that affects our ability to trust is nothing new. However, the sophistication of those technologies (and their capacity of generating truly realistic fake media) and the scale of their reach (and their widespread diffusion) bring the problem and the challenge to a new level. 

Deepfakes thus pose a major threat to businesses, while challenging the current state-of-the-art information theories and technologies as to addressing and mitigating the potential harms they may engender. On the one hand, it makes widely available and cheap to create fake media content, enabling its use in micro-targeting contexts with a high potential of disrupting lives, organizations, and businesses. On the other hand, the techniques by which fake media is generated make it extremely difficult for its detection and mitigation at both social and technological levels. The manipulation of multimodal contents make it exceptionally attuned to the human perceptual system and thus rather persuasive. And, the deep-learning techniques for content generation are increasingly more sophisticated, making it harder and harder to  discriminate automatically between true and false contents. At its core, deepfakes deceive by disguising a harmful action.

It is important to see that the second wave of disinformation is more of a cyber-security problem than a socio-political problem, a key characteristic of the first wave. \emph{Deepfakes} will put in the hands of any cyber-criminal tools to fool and harm people, which can greatly enhance what current phishing and social engineering methods do. For example, with deepfake tools, a false message resembling one's acquaintance voice can be easily created from public samples, and sent through an e-mail or messaging channel, asking for transferring money to what turned out to be a thief's bank account.  Similarly, fake sexual or criminal videos can be created by non-technically minded people with simple, downloadable tools, and be spread in very restricted contexts (such as a WhatsApp closed group, and office channel) to embarrass, bully, or blackmail people.

We follow here the characterization of disinformation proposed by various authors~\cite{jackson2017issue,weedon2017information,marwick2017media,tandoc2018defining,wardle2017fake,vraga2020defining}, who in general look into two different attributes: \emph{falsity} of a specific piece of fake media; and the existence of \emph{intent to harm} from the sender of the information. According to this framework, \emph{disinformation} occurs when both attributes are true, that is \emph{"information that is false and deliberately created to harm a person, social group, organization or country"}. False information which was not created with the intention to harm is called \emph{misinformation}, such as parodies. True information which is intended to harm is called \emph{malinformation}, such as when private information is shared with unauthorized persons. Thus, disinformation is in the intersection of two phenomena, of creating and disseminating false information; and of people and groups who intend to harm other people.

\subsection{The Second Wave of Disinformation}

The current, first wave of fake content and news have created disinformation by relying heavily on the widespread, high-scale sharing through social media networks to achieve society-level negative impacts. We foresee that the second wave of disinformation will be based on much smaller-scale negative effects at a personal level. But, at the same time, it will have greater impact in the aggregate as well as at the individual level by posing greater risks and costs (personal, financial, etc.) to individual people as well as organizations. In other words, the availability of cheap fake media generation tools will create stronger \emph{long-tail} effects where negative impact will be produced by a much larger number of different fake contents, each affecting a smaller number of people. In the limit, there will be cases where a piece of fake media will be created purposefully to attach one single person in a single use, although the attack may be easily replicated to affect millions of people. 

In this context, the defense mechanisms in use today to defend against fake media are not likely to be effective, namely: (i) the manual, chancy detection of fake media by individuals who receive it "by accident" through social media; (ii) the use of human fact-checkers to determine the veracity of a piece of content and the creation of arguments against it; and (iii) the removal of the fake content by the social media platforms to avoid its spread. However, the second wave of disinformation enables the attack of deceptive content in a much more personal (individual) scale, via closed channels on social network platforms or even direct messaging. Both the first and the third mechanisms are likely to be ineffective. At the same time, the sheer number of different fake media enabled by easily accessible deepfake tools will make unfeasible the use of human-based fact-checking.



\subsection{A Prevent and Cure Approach}
\label{a-protect-and-cure-approach}

We propose facing the challenge of the second wave of fake media by taking two different approaches. First, because fake media, as disinformation, is \emph{intended to harm}, it should be addressed as a cyber-security problem, of how to \textbf{protect the users} from effects of exposure to fake media, of social media origin or not. This is an approach much more akin to the framework we use to protect hardware devices from computer viruses than to the current detect-and-remove initiatives from social media companies.


\begin{figure*}[t!]
  \centering
  \includegraphics[width=14cm]{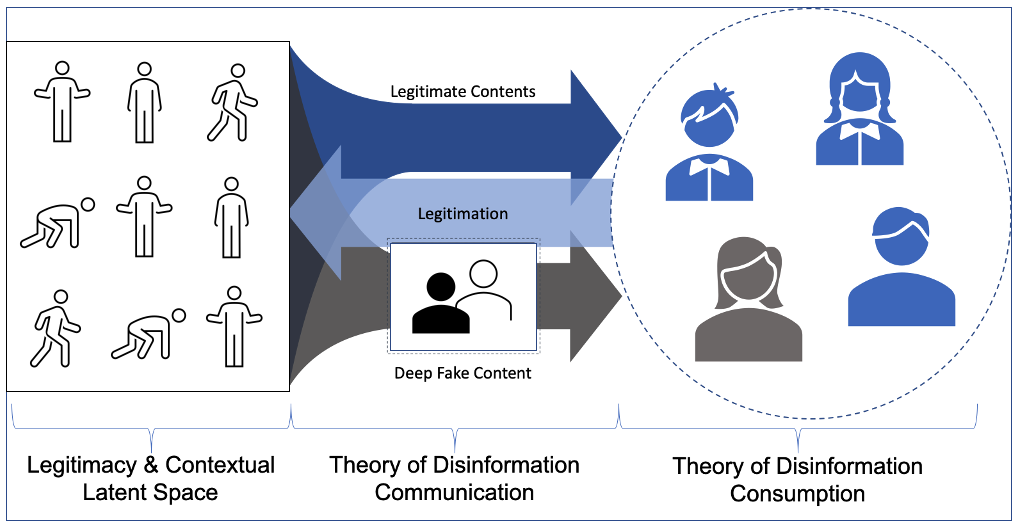}
  \caption{The Science of Disinformation theoretical framework.}
  \label{fig:int-framework}
\end{figure*}

But secondly, and perhaps more importantly, we believe the right approach is not to focus on preventing people from being exposed to fake media (as most of the social media companies are doing), but instead supporting the individual who has engaged with the media and is inclined to or has believed in it. In other words, we want to support the \textbf{prevention and cure of the user} when she is exposed to fake media. In practice, that means having a fake media protector, a piece of software installed on the user's device which can check the media the user is exposed to, and react not only by warning the user of possible issues with that content, but by providing ways and arguments to the user to challenge the veracity of the fake content and verify the truthfulness of the media. In particular, we believe that it is fundamental to help the user after she has already believed in the content of the fake media piece and support her in the process of overcoming the falsity she was exposed to. 

To better illustrate our approach, we will make in the paper an analogy to virus epidemics. 
We understand that, as we write this paper in the midst of the \emph{COVID-19 pandemic}, it could be considered frivolous to use a virus epidemics as a metaphor. In fact, our use is partially opportunistic, since the historical context has made many of the terms known. But more importantly, we believe framing fake media as an epidemic is revealing in many aspects, as noticed by other authors~\cite{kucharski2016study, tornberg2018echo, jin2013epidemiological}.

An important part of the analogy is that it highlights that our approach does not focus on supporting the actors fighting exposure and spreading of virus, such as the CDC in the USA, the WHO, or the governments imposing social distancing and lockdown measures. Instead, we propose to address the effects of fake media at the level of individuals who are or have been exposed to it, like universities and medical research centers do when researching contagion mechanisms and searching for vaccines and remediation drugs. Similarly, to support the individuals who suffered exposure to fake media, we need to find methods and remedies which prevent them from believing in fake media; and which help them recover from the infection caused by the virus, that is, to doubt or stop believing in the received false information.

In fact, this analogy with virus epidemics helps to expose that there are two kinds of remedies to fake media. The first is a \emph{fake media prophylactic}, a preventive measure, for example the use of condoms to prevent HIV. That translates into user practices which helps exposed individuals not to believe in fake media, such as checking the source of the media beforehand. The second is a \emph{fake media antigen}, a curing method, information and engagement strategies to help individuals question the validity of fake media they may have believed in. Examples of the latter are having the user ask for confirmation using alternative channels, from the person she received fake media from; providing counter-factual information from fact-checkers; pointing to inconsistencies in imagery and video; or showing similar, unadulterated media. We believe now is the time to refocus the fight against fake media at finding prophylactics and antigens, or simply, prevention and cure, so it can address the long-tail of the incoming second wave.  



\subsection{A New Science of Disinformation}

To best tackle and address the socio-technical challenges posed by \emph{deepfakes} and fake media in general, we propose to formulate and devise a new science of disinformation as means to help us frame, understand, and address the mechanisms (be social, psychological, or technological) whereby fake media is generated, communicated, consumed, and legitimized as well.

Figure \ref{fig:int-framework} is a high-level representation of our proposed theoretical framework for the new Science of Disinformation. It encompasses two distinct but complementary theoretical frameworks for addressing both the creation of fake media, \emph{the theory of disinformation communication}, and its consumption, \emph{the theory of disinformation consumption}. The former builds on the Shannon-Weaver communication model in order to explore how the sender's intention introduced into the communication process is transformed and communicated as disinformation to the ``receiver''. The latter explores the ways in which people perceive the veracity and legitimacy of such (dis)information and act upon it.

This framework rests on the idea of a common latent space from which existing deep-learning techniques (e.g., GANs) derive their representations and models and, on the other hand, people construct their perceived notion(s) and understanding(s) of an information legitimacy and veracity (that is, from which broader legitimacy context is mentally built). This latent space is both technological (based on statistical or deep learning models of the semantic space) and socio-psychological (based on the people's constructed perception of credibility, trust, and truth). The investigation of the ways in which we could represent and interact with such latent space as means to infer and discern about a message's truthfulness is at the heart of this research endeavor.

One of the main hypotheses of this research is that, at its core, deepfakes undermine people's perception of a message's truthfulness by carrying over the original intent and legitimacy of the latent space from which it is built, while making content modifications that fundamentally transforms its meaning and/or intent. In other words, great legitimacy and contextual meaning are endowed to the message by the very (theoretical and technological) fact that generative models (such as, GANs) builds on the latent space, which is basically a dimensionality reduction operation, for generating new contents. These generative models are theoretically capable of extracting the essence (or the most salient features) that represent an information space. In so doing, they are capable of generating alternative representation of information which carries the latent features (or essence) of the original information space, thus making it extremely difficult to distinguish the real from the fake content. To humans, as the ''receiver'', the generated (fake) message carries the same legitimacy and contextual meaning of original (true) ones. 

The problem is exacerbated by the fact that increasingly, deepfakes focus on generating multimodel contents (based on video and/or audio). In a recent study, it has been shown that multimodel disinformation is considered slightly more credible, irrespective of its source, than textual disinformation \cite{hameleers2020picture}. One may theorize that individuals will resort to``System 1''--a faster pattern recognition--as means to assess the truthfulness of the multimodel message, as opposed to relying on a slower symbolic processing--``System 2'' \cite{kahneman2011thinking}.

This theoretical framing thus provides us the needed tools for establishing a cyber-security ecosystem for automatically detecting and creating prophylactics and antigens for fake media. To be able to create such disinformation cyber-security tools for \emph{the second wave of disinformation}, we face basically three main technological challenges. First, we should be able \textbf{to automatically analyze, detect, and identify fake media} as it arrives in the user's device, preferably prior to the user being exposed to it. Second, we should be able to \textbf{automatically find factual evidence and effective arguments} which can be used in the process of addressing the content veracity. And third, we should create effective, personalized, interactive protocols to \textbf{automatically argue with the user} about the truthfulness of a piece of fake content.

Each of those three challenges are formidable in themselves. As will be further detailed later, detecting and cataloging fake content automatically, at a long-tail scale, is still far beyond current technological capabilities. Moreover, it is a classic \emph{arms race} problem, in the sense that there are active agents, keenly aware of the current detection capabilities, and trying to find ways to circumvent it. 

We affirm here the importance and need of developing a new theoretical framework which can bring strong foundations to the development of effective approaches, such as, new detection algorithms and argumentation technologies. For instance, large-scale machine-based finding and construction of arguments against a piece of content is likely to raise deep technical issues in terms of creating reliable, all purpose fact-checking methods and processes. All this is made more difficult by the fine granularity of the second wave content, which may require access to personal and possibly private information to create effective counter arguments. Although debating technology has improved immensely in the last 10 years, this level of argument and evidence finding is quite beyond today's best systems such as IBM's \emph{Debater}. 

It is also critical to gain an in-depth understanding of the process through which people trust and discredit information so that effective arguing and evidence presentation methods and tools can be developed to support users in recovering from contact with fake media, especially after they believed in it. This is multi-disciplinary challenge, as we will discuss later, which is further complicated by the diversity of people, of the contexts they are in, and in what they already believe.



The scale of those three technological challenges seem to force us into creating new, fundamental principles and ideas which we propose are to be coalesced in a what we call a new \textit{Science of Disinformation}. As discussed, we suggest approaching the construction of this new scientific foundation considering two key theoretical perspective on the theory of disinformation: \textit{communication} and the \textit{consumption}. In what remains, we will explore our proposal in the socio-technical context of the first wave of fake media, and detail further our initial ideas for disinformation theory; then we examine the state-of-art of the current technologies, leading to the outlining of a research proposal. 


\section{Lessons from the first wave of disinformation}



We start by briefly reviewing the major issues and lessons learned from the first wave of fake media to properly establish the socio-technical context of this proposal. The problem of disinformation dissemination has worried several communities from journalists, politicians, academics, and governments for decades, but has become particularly prominent in the last decade in the context of the first wave of disinformation. Focused literature \cite{Vosoughi1146,Grinberg19} shows that fake media spreads farther, faster and deeper on social media than any other type of information (e.g., real stories, terror attacks, and natural disasters). The consequences of such viral dissemination can harm companies and products, eventually having financial consequences, but can also interfere with medical decisions such as jeopardizing vaccination campaigns, disseminating false cures for diseases, or provoking lynching episodes as seen in India \cite{indiaWhats}.

Currently, the COVID-19 pandemic has shown a tidal wave of false narratives that are disseminating faster than the virus itself. The so-called ``infodemic'' has put social technology platforms in constant alert and has made them more proactive in removing hoaxes, fraud, and other abuses. However, combating misinformation and disinformation about the pandemics has proved to be even more difficult in its complexity and scope. The uncertain pandemic scenario makes people more anxious and more vulnerable, increasing their eagerness to stay informed and to engage more with social media as a channel for collective sense making. For many people, they do not intend to spread misinformation. They see sharing information as altruistic because they think that it is helping family and friends to be well informed \cite{peopleFakeNews}. 

At the business level, fake media has also threatened private companies such as the campaign against Starbucks \cite{starbucks} initiated by someone on the online message board ``4 Chan". Some users from that website started creating and spreading fake Starbucks coupons promising free coffee and free access to restroom facilities. Starbucks raced to deny the claims, replying to individuals on Twitter that the information was false. Another incident was during the acquisition of CA Technologies by Broadcom \cite{broadcom}. A fake memorandum apparently signed off by the US Department of Defense warned that the US government would review the transaction for potential national security threats. Because of such fake information, both companies' stock prices went down. These scenarios demonstrate that fake media not only hurts businesses financially, but it creates a toxic atmosphere in which people do not know who or what they can trust. 

\subsection{The Propagation of Fake Media in Social Networks}

Several studies have focused on the dissemination of disinformation, conspiracy theories, and disinformation on diverse online social media platforms, especially for pieces of text disguised as news, the so-called \emph{fake news}. Research on that area focused on limiting the spread online,  on the characterization of such types of content, and on describing bot activity in online social networks. Indeed, these platforms are the main vehicle for public opinion manipulation and fake news dissemination.

For instance, Friggeri et al. \cite{FriggeriAEC14} tracked the propagation of rumors on Facebook and found that fact-checking (e.g., Snopes) affects rumor cascade after a rumor is flagged. Vosoughi et al. \cite{Vosoughi1146} have observed that on Twitter false news diffuses farther, faster, deeper and more broadly than true ones. Resende et al. \cite{WhatsApp19} analyzed the images shared on public groups on WhatsApp during the Brazilian Presidential elections and observed the presence of misinformation among them. Moreover, the work also investigated the propagation of fake images across WhatsApp groups and from/to other web platforms. Finally,  social bots, emulating real users, were used during the presidential election campaigns in the U.S. \cite{Bessi_Ferrara_2016} reaching about 20\% of all posts about the elections; and in France spreading unauthentic documents about a candidate on Twitter \cite{ferrara2017disinformation}.  

Although our research is based on an alternative approach to these investigations, those works show clearly that fake media can very effective in terms both of being accepted as credible and of inflicting harm.


\subsection{The Challenges of Fact Checking}
\label{fact-checking-work-narrative-focus}

Fact-checking is the process of identifying and checking factual statements in text. Traditionally, this task has been done by the journalist community before or after a publication of news articles.
Journalism fact-checking comprises of questioning features such as source, date, location, and motivation to determine the reliability of a fact and must deal with unstructured data, the semantics of the sentences, identification of factual facts as well as the real-time and speed of information spreading. Such activity is still mostly done manually and automatizing such process is not trivial~\cite{Sarr2017AutomationOF}. 

Recently, fact-checking was extended to independent organizations because of the high volume of information being published and the increasing sophistication of political messaging. According to Duke Reporter's lab, there are at least 229 active fact-checking sites in early March 2020 distributed all over the world.  Those agencies daily provide an evaluation of publically-made claims by verifying against authoritative sources and corrections in the form of flagging the falsehoods or providing contextual data. 

Partnerships between social network tech companies and fact-checking agencies have been launched to limit the reach of false posts. For instance, Facebook had used dispute flags to signal fake posts, but that strategy was not efficient. Since not all misinformation is assessed against its accuracy, only a fraction of them will be flagged. Thus, the absence of a warning is ambiguous, meaning either that the post has verified as true or that post was not checked yet. The so-called \textit{implied truth effect} causes a unintentional side-effect in which such untagged posts are then seen as more accurate \cite{Pennycook2017TheIT}. Potential solutions to mitigate misinformation dissemination can go from changing the algorithm by demoting a disputed publication or employing more fact-checkers to deal with social posts.

\subsection{Solutions to Stop the Spread}


The work to stop the spread of fake media has been done largely by fact checking organizations and media sites who have aimed to disseminate the truth and to stop the dissemination of the fake. Examples of current efforts include Twitter and Facebook instrumenting their sites to include warning labels on posts which contain false information~\cite{durkee_2020}. The use of these warning labels has shown to decrease the willingness of users to share fake posts~\cite{mena_2019}.

Other effective methods include removing fake accounts that generate fake media, displaying the credibility ranking of the source, prompting the user with a warning before sharing fake content, and showing the user related, fact-checked articles~\cite{lapowsky_2019}. These methods  aim to make a user less likely to share or believe a given article. Media sites are also implementing methods to lower the chances a user will see false information. Both Google and Twitter do this by prioritizing credible sources when a user searches for a given topic, such as prioritizing the CDC when searching for COVID-19~\cite{romm_2020}. Facebook does this by lowering the ranking of posts in a user's news feed when the post contains false information. Additionally, Facebook will notify those who have read a fake post once that story has been determined to be fake by a fact checking organization~\cite{smith_2017}. 

Another area of research is in increasing the digital media literacy and critical thinking skills of individuals so that if they encounter fake media, they will be less likely to believe it. Teaching the public how to spot fake news has been attempted by researchers from the Cambridge Social Decision-Making Lab. They created a game where players must try to spread fake news and which shows the player what tactics people use in the real world to spread fake news. Their research showed that the game was able to significantly reduce the perceived reliability of fake media~\cite{roozenbeek_linden_nygren_2020}.

\subsection{Shortcomings of Current Solutions}
\label{subsection:shortcomings-cur-solu}

The efforts taken on by media companies listed in the previous section help in slowing the spread of fake news, but fake news is still prevalent because of several reasons. First, there is a blurry line between what is fake media and news, protected-speech satire, and even opinion articles \cite{levi-etal-2019-identifying} which complicates automatic detection and algorithmic suppression.  The effectiveness of other solutions based on increasing digital media literacy or critical thinking of individuals has been questioned to some extent. Danah Boyd~\cite{Boyd2017DidML} argues that large segments of the population distrust experts, media and anyone else who contradicts their world views. In fact, children have been taught to trust information they receive from people they know. Moreover, studies show that repeated statements are easier to process and thus perceived as more truthful than new statements \cite{Pennycook2018PriorEI}. Therefore, once some people are shown a false statement, which agrees with past information, it is hard to convince them otherwise even when presented with the truth from a credible source.  Indeed, disinformation deals with psychological domains such as the individual social identity as well as their sense of group belonging. 

Another point is that social platforms have not provided mechanisms for companies to report fake media before it potentially affects a share price or sales. Companies themselves have to be vigilant about everything that is posted about them in social media. Timing is crucial during social media firestorms and crisis management teams do not seem to be efficient enough to deal with such unmatched power. 

The time it takes to determine that a given media is fake further complicates the problem. Currently, the process to determine what is fake is very manually intensive. In the most recent epidemic of fake news about COVID-19, fact-checkers from all over the globe coordinated with the International Fact-Checking Organization. They communicated over Slack and Google Sheets to share their fact-checks so that they could then translate and republish articles that contained the truth~\cite{poynter_2020}. This example demonstrates the multiple, time-consuming steps necessary for fact-checking even at the small scales of the first wave of disinformation. Also, during the fact-checking investigation period, fake news is able to propagate and "infect" individuals, and once an individual becomes infected, there is no way to ensure that they will ever be exposed to the truth or that they will believe the truth.

While others' recent efforts target understanding and limiting the spread of false information, our research presents an alternative approach to these investigations which focuses on the information security of the individual. Since we cannot entirely stop fake media from being created and spreading, we aim to ``cure" those who have become ``infected" and understand the psychology that causes one to share false information. To do this, we look to understand the user's perception of a given piece of information to better understand their beliefs and motives. Our solution is independent from how fake media disseminates and is aimed at creating personalized cures which will help change an individual from believing in fake media, to at least raise doubts about it, and, hopefully, open him/her to believe in the truth. To achieve this, we propose first to rethink key elements of information theory to model appropriately the falsity and the intent to harm of fake media, as detailed in the next section.








\section{The Theory of Disinformation Communication}
\label{section:disinformation-theory}

Disinformation is false information that is spread with a harmful intention, e.g., to deliberately deceive the information receiver. To study disinformation, we need to understand its two critical components: false information and harmful intention. Inspired by Shannon's information theory~\cite{shannon2001mathematical}, we propose to create a \emph{theory of disinformation communication} which fundamentally models how disinformation communicates between individuals.

The Shannon-Weaver communication model characterizes communication as a system process, where a message is transmitted from an information sender through a noisy channel and received by an information receiver. Built on top of this fundamental model, we explicitly introduce \emph{intention} into the communication process to characterize how disinformation is transmitted between individuals. Intention, as a fundamental mental state,  describes behaviors in individuals who have desires and attempt to accomplish goals that are directed by beliefs. In disinformation, the sender has a harmful intention to influence the receiver's opinion.

In the \emph{Theory of Meaning}~\cite{grice1975logic}, Grice introduced the concept of \emph{non-natural meaning}, which describes the intentions of the speaker in communicating something to the listener. According to Grice, the intention of a speaker $A$ by communicating $x$ is roughly equivalent to ``A uttered $x$ with the intention of inducing a belief by means of the recognition of this intention''. We may potentially leverage Grice's theory to help us model the communication of disinformation as well as how to block such process, i.e., combating disinformation. Note that Grice's Theory of Meaning is one candidate theory that may help us better understand intention in information communication. We are also investigating other research work on intention in philosophy and psychology. 

\begin{figure*}[t!]
  \centering
  \includegraphics[width=14cm]{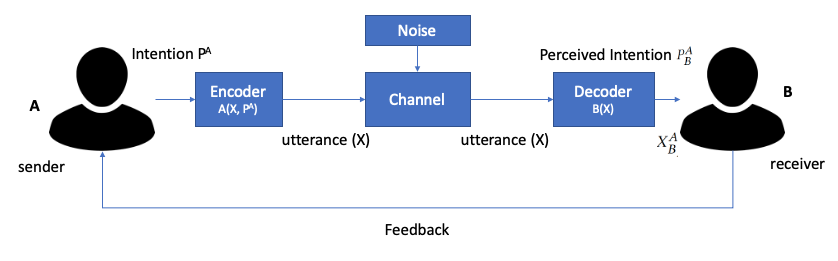}
  \caption{The Grice + Shannon-Weaver Communication Model.}
  \label{fig:gsw-model}
\end{figure*}

We introduce Grice's Theory of Meaning into the Shannon-Weaver communication model to describe how disinformation is communicated between individuals. We decouple disinformation into intention and false information. Figure \ref{fig:gsw-model} shows the communication process, where $A$ is the information sender, $B$ is the information receiver, $X$ is the false message, and $P$ is the intention of an individual. The intention $P$ can be a latent variable or vector. Let $P^A$ denote the intention of $A$ and $P^A_B$ denote $B$'s perceived intention of $A$. The encoder of $A$ converts $A$'s false message $X$ and the intention $P^A$ into an utterance of $X$, that is, $A(X, P^A) = utterance(X)$. Therefore, an utterance comprises both the message and the intention. After $B$ receives $utterance(X)$, $B$ decodes the  $utterance(X)$ into a message $X^A_B$ and a perceived intention of $A$, i.e., $P^A_B$. We use $X^A_B$ to denote the message which $B$ receives from $A$. Note that the encoder and decoder in Figure \ref{fig:gsw-model} are more general than those defined in the original Shannon-Weaver communication model, which are only used to convert between message and transmission signals.

This new definition may help us recognize that the transformation of a particular input is also influenced by its semantic context and personal belief. Suppose $P$ is a latent vector. We can use the cosine similarity $cos(P^C, P^A_B)$ to indicate the degree that $B$ recognizes the intention of $A$, where $1.0$ indicates that $B$ fully recognizes $C$'s intention, and $0$ indicates that $B$ fails to recognize $C$'s intention and is fully deceived by $C$. 

Given $B$'s perceived intention, an additional feedback mechanism can be introduced to further query and clarify $A$'s encoded intention to ensure that the message was well received. However, the feedback between sender $A$ and receiver $B$ may not always be feasible, as for example in the case where sender $A$ has posted a video and $B$ has no direct access to the original content posted by $A$. If there is a direct connection or proximity between the sender and receiver, then a feedback loop may exist between both parties. Otherwise, a 3rd party or external adversary $C$ may have access to the feedback, or perhaps no feedback exists.

\begin{figure*}[t!]
  \centering
  \includegraphics[width=14cm]{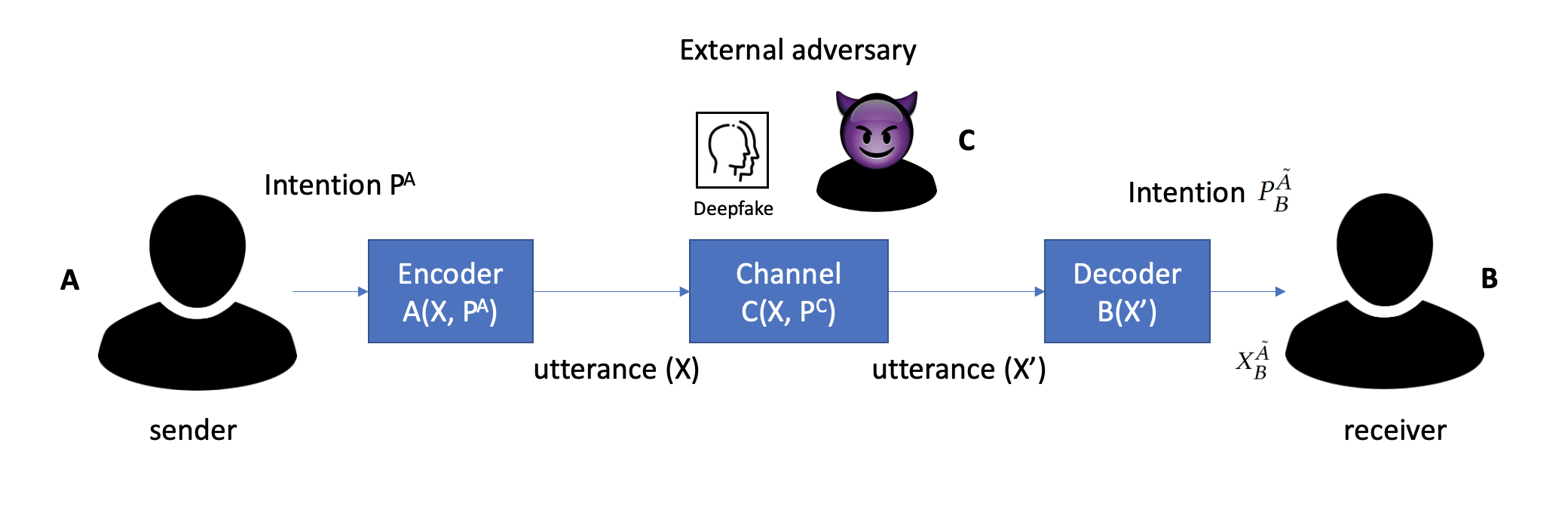}
  \caption{A deepfake communication example based on Disinformation Theory.}
  \label{fig:df-model}
\end{figure*}

We can now use the proposed model to describe the communication of disinformation. Figure \ref{fig:df-model} describes such an example, where an external adversary individual $C$ with a harmful intention sits in between the communication from $A$ to $B$. Suppose public figure $A$ presents a video statement $X$ with a particular intent $P^A$. The goal of the external adversary $C$ is to manipulate the $utterance(X)$ from A and deceive the receiver $B$ with a harmful intention $P^C$. $C$ changes the original message from $X$ to $X'$ using a DeepFake model, and generates a new $utterance(X')$ through their encoder, i.e., $C(X, P^C) = utterance (X')$. After $B$ receives $utterance(X')$, $B$ uses their decoder to interpret the message $X^{\tilde{A}}_B$ and intention. We use $\tilde{A}$ to indicate the case that the information seems to be sent by $A$, but actually not. If $C$ is an unknown adversary to $B$, the perceived intention $B$ is $P^{\tilde{A}}_B$. Otherwise, if $B$ knows that the received $utterance(X')$ is directly from $C$, $B$'s perceived intention is then $P^C_B$. $B$ would become ``dis-informed'' if $B$ does not recognize the harmful intention. We can also use this model to describe cases where $A$ directly communicates disinformation to $B$ or both $A$ and $C$ are adversaries. This model therefore allows us to consider different attack scenarios and provides an approach to describe the underlying mechanisms of disinformation communication. 

While in this work our focus is on disinformation communication between individuals, we can further extend it to a network of communications where disinformation propagates. Leveraging methods from network science and graph theory, we can model a much more macroscopic perspective of how disinformation is communicated and evolved. This analysis can help us have a global network view of the dis-informed status of individuals.

\section{The Theory of Disinformation Consumption}

In the previous section we propose a new theoretical framework, i.e. \emph{theory of disinformation communication}, inspired by Shannon's information theory and built on Grice's theory of meaning to help us understand and address the problem of how disinformation transforms the original intention of a message as means to deceive the receiver. In so doing, our goal is to devise effective automated tools for detecting fake media by ways of blocking the communication of disinformation. 

However, to attain a fuller picture of the mechanisms by which disinformation affects and ultimately impairs human judgments, we introduce a distinct but complementary theoretical framework, the \emph{theory of disinformation consumption}, which attempts to frame the ways in which human actors (the ``receiver'') make sense, make decisions, and take actions upon the information at hand. Ultimately, we are concerned with devising a new socio-technical approach which can influence and maybe prevent people from making decisions and taking actions, often inadvertently, that could harm themselves, their organizations, and others.

We resort to a question of meaning making and decision making in which their outcomes can be perceived and articulated by the decision-maker (e.g., the perceived and potential benefits and risks of taking an experimental drug or the potential harm of inadequately sending one's personal information to a stranger). In other words, the majority of the studies on the spread and use of fake news investigates the effect of misinformation and disinformation as an action whose potential outcomes and impacts are not (or cannot be) fully apprehended or anticipated, such as, the spread of political rumours, or the dissemination of vaccine misinformation \cite{wang2019systematic,poland2010fear}. Instead, in this research, we are concerned with a particular problem in which the role of disinformation is to attempt deceiving people (or organizations) to take actions that they would otherwise be able to perceive its harmfulness and not to carry them out, should they know its falsity. 

We thus contend that, contrary to the typical cases of the first wave of disinformation, where the outcomes are often perceived individually as of low impact or cost (\emph{``after all, it may cause no harm to share this point of view for the treatment of COVID-19 that has no scientific basis but seems rather logical,"} one may think), the outcomes (harmful or otherwise) of micro-targeting individuals and organizations with disinformation, which comprises the second wave of disinformation, are of clear impact to the receiver, when its falsity is unveiled. Table \ref{table:first-sec-comp} presents potential impact of one's action or decision pertaining to a potential disinformation content, such as, further sharing it or acting upon it, per perceived attributes, such as perceived costs, benefits, and risks of carrying out certain action and the potential impacts to the individual and the community at large. 

\begin{center}
 \begin{table}
  \begin{tabular}{ l c c } 
   \hline
   \textbf{Attributes} & \textbf{First Wave} & \textbf{Second Wave} \\ [0.5ex] 
   \hline
    Costs & Low & High \\ 
   \hline
    Benefits & Low & High \\
   \hline
    Perceived Risks & Low & Low \\
   \hline
    Individual Impact & Low & High \\
   \hline
    Collective Impact & High & Low \\  
   \hline
  \end{tabular}
  \caption{Attribute comparison of first and second waves of disinformation.}
  \label{table:first-sec-comp}
 \end{table}
\end{center}

The search for a socio-technical solution certainly requires an in-depth understanding of how humans interact with, perceive, evaluate, judge, and respond to the truthfulness (or falsity, for that matter) of a piece of information. Undoubtedly, the processes by which humans construct truthfulness are sociologically, psychologically, and philosophically complex. In short, they rely not only on one's understanding of the information at hand, but the ways in which it fits into a broader socio-cultural context as well as one's own belief system(s) and intentions. 

The perception of an information truthfulness or falsity can be thought of as the result of one's ability of building social consensus (\emph{``do others also believe it?''}), finding evidence to support it (\emph{``is there enough supporting evidence(s)''}), meeting one's internal and prior consistency (\emph{``is it compatible with my prior knowledge and experiences?''}), building internal coherence (\emph{``does it sound `reasonable'? does it tell a `reasonable' story?''}), and credibility (\emph{``does it come from a 'credible' source?''}) \cite{schwarz2016making}. Of course, people do not take these elements as categorical criteria they ought to go systematically through as part of a mental check-list when assessing the truthfulness or falsity of a message. In practical terms, people often conflate them into one or more broader perceptual categories, such as, the sender credibility, the message legitimacy, and the message intention, when judging its truthfulness or falsity.  

By taking this practical perspective on the ways in which people judge the intention of a message, we build \emph{the theory of disinformation consumption}. People are constantly foraging the information landscape \cite{pirolli1999information} as they navigate through the plethora of information to which they are exposed. As a result, broader and generalized categories (or logical classes) of things emerge (e.g., \emph{``all politicians are corrupt,''} \emph{``COVID-19 is dangerous,''} \emph{``the Government is forcing me to stay at home,''} and the like) as they attempt to make sense of the realities based on the information ecology with which they engage. Consequently, these categories shape how they judge any new piece of information's truthfulness as measured by its perceived coherence with respect to their constructed mental categories. This could be thought of as mechanisms for maintaining their prior mental structures consistent. In this process, the ``sender's'' credibility and legitimacy are also taken into consideration. As people organize and incorporate new pieces of information into those categories, they start adjusting as well as reifying their personal beliefs and biases. In addition, they make use of these as reference points against which they assess and make decisions based on a piece of information at hand. 

\begin{figure*}[t!]
  \centering
  \includegraphics[width=14cm]{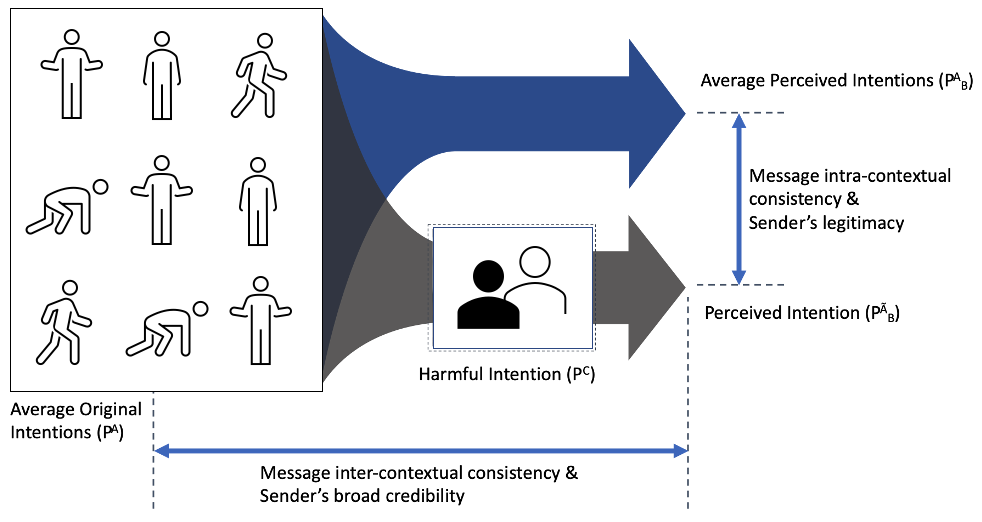}
  \caption{Two theoretical distances between an interpreted message and the latent space: inter- and intra-contextual analysis.}
  \label{fig:intra-inter}
\end{figure*}

Figure \ref{fig:intra-inter} borrows the original notion of constructed latent space (cognitively or otherwise) as the reference against which people measure the consistency, legitimacy, and credibility ``distances.'' It conceptually depicts two basic distances: (1) between the disinformation utterance's perceived intention ($P^{\tilde{A}}_B$) and the aggregated flow of similar perceived intentions by the original sender (Avg $P^A_B$); and (2) between the disinformation utterance's perceived intention ($P^{\tilde{A}}_B$) and the aggregated space of original intentions (Avg $P^A$). These are two conceptual distances but they help us formalize and address the research questions of automatically creating arguments for making the ``receiver'' question the truthfulness of the received utterance, thus refraining from taking the wrong action or making the wrong decision.

Based on this theoretical framework, we aim to investigate three critical elements that comprise: what we are calling, a \emph{disinformation ecosystem}, a complex system of actors, parts, and relationships, and how they come together to enable or hinder the impact of disinformation on individuals, organizations, and society. As we move forward in this research endeavor, we will investigate the mechanisms and processes by which people consume (dis)information ---\emph{disinformation consumption}--- as a means to allow us to create counter-measures, such as, fake media antigens (see Section \ref{subsection:research-antigen-gen}) and prophylactics (see Section \ref{subsubsection:gamification}).

We will also investigate the processes by which people make decisions in the context of a disinformation threat ---\emph{disinformation economy}. Traditionally, studies on decision-making, from psychology, behavioral economics, and operations research, rely on the notion of decision-making under uncertainty, where, broadly speaking, the main problem is the lack of available information to make the most accurate decision. In contrast, in the second-wave of disinformation, the uncertainty of decision-making does not rest on the lack of information, but on one's ability to discern what is fake or not. And, finally, we will explore argumentation mechanisms that challenge people's perception of truth and intention, just enough to make them question the truthfulness of the utterance--\emph{disinformation arguing}. We aim at creating automatic mechanisms for creating arguments that would prevent people from inadvertently making a decision that may cause harm to themselves, to their organizations, and to others. To this end, we plan to investigate the complex network of actors and information that come to influence one's perception of truth or not when making a decision.




\section{State-of-art in fake media generation and detection}

Having looked to the fake media problem, both in terms of communication and consumption, from a theoretical point of view, it is also important to explore them in terms of the current technology which enables and delivers them.
In this section, we describe the fundamental frameworks and algorithms behind many of these generated fake media, describe the components for fake media generation, and review some of the state-of-the-art manual and automatic conventional and deep learning methods and tools commonly used to detect deep fake generated media.

Recent advancements in deep learning models as well as the abundance of high quality datasets, powerful computation systems, and large storage capacities has enabled researchers to develop high quality media. Deepfakes are a class of techniques which synthesize or augment facial features of an image based on some attributes, and come from a class of deep learning algorithms known as \textit{Generative Adversarial Networks (GANs)}. These techniques can be applied laterally across any media domains, as long as there are sufficient samples that can replicate and synthesize complex features exhibited within the given distribution of the dataset. For instance, some methods can not only enable transfer or mapping of attributes for image-to-image tasks, but also from other domains such as text-to-image or speech-to-video.

The media content generated by these models have paved the way to the generation of not only professional and realistic-looking media forgeries, but also to the wide applicability of these models to alter media content for nefarious intentions \cite{chesney2019deepfakes}.
Detection tools and methods have been developed over the years to be able to detect fake media content and can be divided into two general categories, (1) conventional detection methods that look for artifacts in manipulated data and employ supervised machine learning methods with hand-crafted features, and (2) deep learning methods that learn features directly from the data.

\subsection{Framework for Fake Media Generation}
Deepfakes are generated by deep autoencoders and Generative Adversarial Networks. The first attempt of deepfake generation was based on a paired autoencoder-decoder network. The autoencoder extracts latent features of face images (e.g., latent faces), and passes them to a decoder, which reconstructs the face images as much as possible. The performance of the network is measured by how well it reconstructs the original image from latent faces. Different loss functions can be implemented as reconstruction objectives during the training phase (e.g., \textit{Faceswap} uses the mean absolute error (MAE) loss as the objective function). The pair of autoencoder-decoders is used to reconstruct a source image set and a target image set separately. 
In order to generate deepfakes, (1) a latent face generated from a source image is passed to the decoder for the target image set, (2) the decoder will then try to reconstruct a target image from the input latent face of a source image, and (3) if the autoencoder-decoder network is trained well enough, the latent faces can represent critical facial features of original images. The finally generated target image will ``contain" critical facial features of the source image. 

The key to advanced deepfake generation technologies relies on improving the  performance of the encoding and decoding phases. \emph{Faceswap-GAN} introduces not only an adversarial loss reconstruction objective function in the autoencoder-decoder architecture \cite{shaoanlu_2019}, but also adds a perceptual loss implemented by \emph{VGGFace} to the objective function. The perceptual loss improves direction of eyeballs to be more realistic and consistent with input face, and it smooths out artifacts in the segmentation mask, resulting in higher image output quality. Other traditional facial expression manipulators include \emph{Face2Face} \cite{thies2016face2face} and \emph{NeuralTextures} \cite{thies2019deferred}, which transfer facial expressions from one person to another in real time in such a way that the identity of the other person is not obscured. Likewise, \emph{StarGAN} ~\cite{choi2018stargan} is able to learn mappings among multiple domains, and able to translate an image into a corresponding domain. An image can be modified to change the expressions to angry, contemptuous, disgusted, fearful, happy, among others. 

\subsection{Components for Fake Media Generation}
Many of the state-of-the art models show that depending on how learning problems are framed, various methods and techniques to generate fake media are possible. The key to formulating a learning problem lies within two fundamental components: (1) content synthesis vs. content augmentation, and (2) domain reference mapping. Many of these generative models can either fall between two different categories: synthesis-based or augmentative-based. Whereas \textit{synthesis} is based on a principle where we construct the final output of the generated image based on some random input, such as some uniform noise distribution, \textit{augmentation} pertains to the modification of the input to generate a new composition such that the general feature structure of the input still holds, but with some discrepancies from the original source.

\subsection{Fake Media Detection}

The field of \emph{Multimedia Forensics} has provided investigators and scholars with a set of tools, methodologies, and frameworks to identify and detect the augmentation of digital content. The field is comprised of two domains, \textit{source identification} and \textit{forgery detection}. \textit{Source identification} is concerned with the discovery of the origin of how and where content originated from, while \textit{forgery detection} provides a method for assessing the integrity and authenticity of media content.

Some of the methods used for both domains are manual and rely on human intuition to determine if the content seems unnatural, while others rely on using automatic detection techniques. Fake media detection models have typically fallen into detection techniques which in general include experts fact-checking the content, crowd-sourcing approaches to do a collective consensus, and various machine learning, natural language processing, and hybrid approaches~\cite{nguyen2011intelligent}. 
We review next some state-of-the-art tools, conventional methods, and deep learning methods for the manual and automatic detection for the following types of media: image, video, audio, and text.

\subsubsection{Manual detection}
\label{manual-detection}

The \textit{image-based} manual detection methods typically look for a violation of physical laws, such as shadows not appearing as expected given the position of the light source, or use reverse image search to see if the photo originated from a different context. Techniques based on error level analysis transform the original image into a gray-scale image in order to highlight the intensity of the light source and the texture seen in the photo. The resulting photo is then interpreted by a human to determine if part of it has come from a different source or has been modified ~\cite{*_2020}. Even with these techniques, it has been shown that humans are significantly worse than machines at detecting fake images ~\cite{schetinger_oliveira_silva_carvalho_2017}. 

\textit{Video-based} manual detection might use the same tactics as the ones used to detect fake images. In order to detect fake videos, one can look for signs of irregular eye blinking or objects/facial features that appear or disappear in an instant. While these tactics have shown some promise, the creation and detection of fake videos is an arms race and at scale, thus human detection of fake videos becomes unfeasible. For instance, Youtube has resorted only to add fact-ckecks to videos which are produced from reputable media publishers, but \emph{''it won't be fact-checking individual videos ---a task that would require a gargantuan effort and an uncountable number of human hours to accomplish''}~\cite{ajdellinger_2019}.

\textit{Audio-based} media checking focus on assessing how a voice sounds to the human ear. This means that the timing between sentences flows normally, the pitch of the voice does not change drastically, the voice does not sound ``robotic'', and that there are not mispronunciations. A study conducted in 2015 found that humans were able to detect fake audio 50\% of the time ~\cite{10.1007/978-3-319-24177-7_30}. However, due to the increased performance in generating fake audio, experts say that it is now ``extremely difficult for the human ear to discern a real voice from a deepfake''~\cite{sharkey_2019}.

\textit{Text-based} media validation is done by breaking a given text into individual claims, and identifying each source and evidence that supports it. If sufficient evidence is not provided, the investigator will search online and offline for credible sources that confirm or refute the presented claim. Every potential source is evaluated by author/publisher to establish credibility, determine the author's point of view to spot for any potential biases, and observe the date of publication to ensure that it is relevant to the claim in question ~\cite{research_guides_fordham_2020}. Experts in the field might be contacted to better evaluate the information and reach a final verdict about the claims.

\subsubsection{Automatic detection}
\label{sota-auto-detection}

The \textit{image-based} detection systems have used many traditional classifiers and deep learning methods. Deepfake generators make it very difficult to detect fake images since these models (1) learn complex distributions from training data sets, and (2) generate realistic and high-quality data very similar to the training data. Conventional detection methods include blind-based methods or model-based methods that look for specific artifacts or imperfections generated by the camera hardware and software components (e.g., lens, sensor, demosaicing algorithm) ~\cite{verdoliva2020media, gallagher2008image, cozzolino2014image, chen2011detecting}. Conventional methods have worked well in most general cases but if images are re-sized or compressed, or the train and test data sets do not come from the same distribution, the performance of these detectors drops significantly. 
Deep learning detection methods such as Convolutional Neural Networks (CNN) have been employed to combat these adversarial networks. Hsu \textit{et al.} \cite{hsu2020deep}, propose a two-phase detection method in which a common fake feature network (CFFN) is used to extract features that are introduced to a CNN that is concatenated to the last layer of the CFFN to detect fake images. Some features that have been widely used for detection include face warping artifacts, facial expressions and pose visual artifacts, as well as color and image-related feature discrepancies. While some deep learning models require very little pre-processing, large quantity of images might be required to train these models so they generalize well.

\textit{Video-based} detection systems can be categorized into methods which process the temporal features across a set of video frames, and models which process discriminant features captured from single frames. Methods that process temporal features across video frames typically follow a (1) pre-processing phase to detect, crop, and align the face regions, and (2) a detection phase to process all the frames in that sequence by combining CNNs and a recurrent neural network (RNN) architecture such as Long Short Term Memory (LSTM) or Gated Recurrent Units (GRU). The CNN component extracts frame-level features, whereas the neural network looks at the temporal sequence. Some of the recent works include a detection model that proposes a CNN followed by a RNN network to capture temporal inconsistencies \cite{guera2018deepfake}. The video sequence is first fed to a CNN to extract features from each frame, features are concatenated into a single feature vector, and then forwarded to a LSTM sub-network for processing and classification. Another similar work, \cite{sabir2019recurrent}, exploits the use of a recurrent convolutional model to analyze temporal artifacts as a way to indicate if faces in a video sequence are abnormal or not. Manipulation tools do not enforce temporal coherence during the synthesis process, thus this weakness can be exploited. 

Regarding the methods which process visual artifacts within frames, these methods typically decompose videos into individual frames to obtain the discriminant features that capture some visual artifacts. These methods are typically deep (e.g., CNNs with multiple hidden layers to extract finer features) or shallow classifiers (e.g., Support Vector Machine (SVM) and neural networks with one layer). Some of the recent works include the use of capsule networks, where latent features are extracted using the VGG-19 network and fed into a network of 3 capsules (dynamic routing between capsules algorithm is employed to boost detection performance) \cite{nguyen2019capsule}; the use of optical flow fields to identify inter-frame dissimilarities/correlations and be used by CNN classifiers \cite{amerini2019deepfake}; the extraction of biological signals hidden in videos, \textit{FakeCatcher} \cite{ciftci2019fakecatcher}; the use of the Scale-Invariant Feature Transform (SIFT) algorithm to analyze successive frames and differentiate between real and fake videos \cite{djordjevic2019deepfake}; and the detection of the lack of eye blinking in videos to exposed synthesized videos \cite{li2018ictu}. 

The advancement of deep learning-based models such as GANs has led to the propagation of high-quality fake video content. There is a large collection of fake video detection systems and due to their complex architectures, they may not generalize well as more sophisticated fake content generators evolve. Although one may be inclined to use image detection models, these models might not work well for video content since (1) videos have spatial and temporal characteristics that describe the media content rather than just spatial content as found in static images, and (2) video compression degrades the video data to a point that it may be difficult to assess due to the affine transformations (rotation, scaling, translation) needed to properly pre-process the video content in some instances.

\textit{Audio-based} detection methods focus on classifying the signal as original or synthetic ~\cite{alegre2013one, campbell2006support, villalba2015spoofing}. Synthetic speech detection consists of (1) determining the relevant features that capture the signatures of the signal that characterize the artifacts created during the synthesis phase, and (2) using a model to represent those features and determine if the audio is real or not. Feature extraction techniques for synthetic speech detection have focused on features such as the logarithm of the power spectrum, the power spectral flux of consecutive frames, the modulation spectrum of the signal, and many others \cite{Sahidullah2015ACO}. Probabilistic and supervised machine learning models such as Gaussian Mixture Models (GMM) and SVMs have been used to classify the signal as synthetic or original \cite{alegre2013one, campbell2006support, villalba2015spoofing}. In spite of the many approaches that require feature engineering and used of various machine learning models, speech synthesis technology has become more sophisticated in recent years due to the use of deep models such as WaveNet, a generative model for raw audio \cite{oord2016wavenet}, or \textit{RealTalk}, a system that consists of multiple deep learning systems to generate life-like speech using only text inputs \cite{dessa_2019}. In order to deal with audio fake content, some audio detection models format audio clips into spectrograms, heat-map like visual representations of the sound waveforms, and use them in models with temporal convolutional layers to predict if the raw audio is either real or fake \cite{dessa-oss_2020}.

\textit{Text-based} media detection methods are typically based on text generative models such as \textit{Grover}~\cite{zellers2019defending}, which generates a news article given an original article components (e.g., domain source, headline, article content). The fast improvement of language models has led to the development of tools such as Giant Language model Test Room (GLTR) \cite{gehrmann2019gltr} to visually help and teach humans in deciding whether some of the text was generated by a model or not. The idea is to detect generated text via probability distributions by using a similar model that was originally used to generate the text. Given a sentence, the model would check if that text was generated by a language model such as GPT-2 \cite{radford2019language}.

\section{State-of-Art in Machine Argumentation}
\label{section:soa-arg-gen}

Having reviewed the state-of-art of fake media generation and detection, we review now the technologies which can be used to prevent and cure individuals exposed to fake media. Once an individual has read and believed a piece of fake media, effort is required to convince that individual that what they believe to be true is actually false. 

Helping individuals to understand that a piece of information is false and/or intended to harm him/her is a difficult task. First, it takes a significant amount of time to create arguments or to create fact checks that can be used to convince an individual of the truth. Second, it is difficult to change an individual's beliefs when that individual was initially exposed to false information, and it becomes more difficult to change their mind the longer they believe the false information.

For these reasons, it is important to challenge the individual quickly, as soon as he/she has been exposed to potentially false information. To accomplish this, we need to have automatic generation of arguments and/or explanations which may allow an individual to question the veracity of information contained in video, image, or text. 

In this section we explore the state-of-art of the techniques in automatic argument generation to see which techniques may be used to support questioning the veracity of fake media.

\subsection{Overview of Argument Generators}
There are two classes of argument generators, ones that select their claims from a text corpus and ones that learn to generate their own claims. In this paper we focus on argument generators that rely on a text corpus since they have been proven to be used effectively in practice, for instance, IBM's \emph{Debater} \cite{slonim2018project}. Here we do not go into details, but outline the steps taken by most argument generators when creating a sophisticated argument so that we can see how the overall process is related to fake media argument generation.

First, argument generators rely on a large corpus of text which they search through to detect claims and evidence about the topic in question. In the context of fake media, the claims would be found from media surrounding the topic and results from detection techniques that determine the validity of a claim. Next, an algorithm is used to select the strongest claims from all that were found. These claims are then organized by themes so that a cohesive narrative for the argument can be made.

When in a setting where there are two "players" involved in the argument such as in a debate with an opponent, or in the context of fake media with a disinformed user, the argument generators must listen to the response of the other and create a rebuttal. Creating a rebuttal involves detecting the other's arguments and then again finding claims and evidence to create an argument in response. This is the same general process that will be required in creating arguments in the context of fake media. We next highlight the nuances. 

\subsection{Challenges for Argument Generation in Fake Media}
\label{subsection:nuances-arg-gen}
In the context of fake media, current argument generation techniques must be extended to include several factors. First, we must augment text corpora with claims given from detection results. This way, generated arguments can include explanations and evidence for why a given piece of media is fake. Additionally the intent of an individual who is posting content differs from normal debates where the intent is to convince you that one stance is better than the other. With fake media, a malicious user can have intents such as political persuasion or financial gain. This information about the intent of a user can be modeled and then included to create a more convincing argument. 

Our research directions outlined in section \ref{subsection:research-antigen-gen} show that we can extend the capabilities of current argument generation techniques by judging quality of arguments based on their effectiveness to change someone's perspective and by using psychological frameworks to generate arguments and that provide a new metric for argument quality.

\subsection{Determining the Quality of Arguments}
One challenging problem of argument generation is to rank the quality of a claim so that the most convincing claims can be used to create a narrative that is able to persuade an individual. In 2020 a survey was conducted on argument generation, and one aspect it reviewed was the creation of argument data sets \cite{lawrence_reed_2020}. The proposed techniques most often require an annotator to determine the quality of an argument. There are different scoring mechanisms which can be used to select a score for the quality of an argument such as annotators ranking an argument as convincing or how an annotator ranks an argument compared to another argument. 

The IBM Debater team recently released a data set which uses a new method that determines the quality of an argument that considers the credibility of the annotator \cite{gretz_shai_2019}. In a few instances, claims were classified by which type of sentence the claim could function as in an argument, such as method, result, or conclusion sentences. These data sets can be classified by an unsupervised approach that looks for keywords such as "this conclusion..." \cite{arg-mining, houngbo_mercer_2014}. The problem of determining the quality of a claim becomes more challenging when there are multiple sources that present conflicting claims. This area of research is very relevant to fake media where there are often conflicting claims. There are automated methods to help determine the credibility of a source and the truthfulness of a claim. One solution to this problem that has surpassed performance on previous state-of-art techniques relies on probabilistic soft logic \cite{samadi_talukdar_veloso_blum}.

None of the data sets we have found for argument generation look to determine the quality of an argument based on the argument's ability to change a person's mind. They all rely on annotators to label the quality of an argument based on how convincing they believe the argument to be. In section \ref{subsubsection:research-arg-eff} we discuss how we plan to introduce new data sets that determine quality based on the effectiveness of an argument to change one's mind.

\section{A Research Proposal}

As described in Section \ref{a-protect-and-cure-approach}, we propose that fake media should be addressed as a cybersecurity problem where fake media prophylactics and antigens are created, automatically and in a principled way, to protect users from the harmful effects of fake media exposure. This means having a fake media defense system which can check the media a user is exposed to, and in doing so, provide ways and arguments to challenge the veracity of the content in order to help the user throughout the process of avoiding or overcoming its falsity.

In this section, we outline a research proposal and some of the areas we would like to further investigate, and how they can lead to the creation of an effective ecosystem of prevention and cure. The research proposal is composed of six main themes:
\begin{enumerate}
    \item Creating the theory of disinformation communication;
    \item Creating the theory of disinformation consumption;
    \item Developing automatic fake media detection methods;
    \item Developing fake media antigen creation systems;
    \item Establishing fake media prophylactics methods;
    \item Addressing ethical questions and issues.
\end{enumerate}

Themes 1 and 2 are theoretical in nature, while themes 3 to 5 are mostly technology-focused. Since dealing with falsity, intent to harm, and media monitoring and argumentation systems certainly raise many kinds of ethical questions, we propose theme 6 to be sure to address those issues from the beginning. In the next sub-sections, we explore some ideas and aspects of this research proposal and how we plan to conduct research in each of those theme.

As part of this proposal, we would like to engage with global networks of researchers and organizations (e.g., \emph{DeepTrust Alliance}) interested in sharing theoretical results, datasets, scientific results, methodologies, and code to fight against disinformation, as well as to establish and participate in public competitions to detect deepfakes and fake media. 

\subsection{Creating the Theory of Disinformation Communication}
\label{subsection:create-theory-dis-communication}

With the proposed theory of disinformation communication, we can investigate how to protect individuals from disinformation by producing two substances: a curing substance, the \textit{antigen}, and a preventative substance, the \textit{prophylactic}. Let us use the deepfake communication in Figure~\ref{fig:df-model} as an example. Our main objective is to maximize $cos(P^C, P^A_B)$, that is, to enable the information receiver $B$ to best recognize the intention of a disinformation sender $C$. Therefore, the objective function is: $$max\hspace{0.1cm} E{[cos(P^C, P^{\tilde{A}}_B)]}.$$ 

This can be studied in the context of the following two research problems. 

\subsubsection{Fake media antigens}\label{plan_antigen}

Suppose the information receiver $B$ was ``infected'' by the fake media, that is, $B$ failed to recognize the intention of the fake media sender $C$. We have $cos(P^C, P^{\tilde{A}}_B) \approx 0$. The objective of producing fake media antigen is to provide a set of arguments $Y$, which challenge $B$'s received information $X^{\tilde{A}}_B$. As a result, $B$ can improve their recognition of $C$'s intention by analyzing the arguments. 
That is, the value of $cos(P^C, P^{\tilde{A}}_B)$ is increased. We thus have the following optimization problem: 
$$argmax_{Y} \hspace{0.1cm} E{[cos(P^C, P^{\tilde{A}}_B)]},$$
where the objective is to choose a set of arguments $Y$ so that we can maximize the similarity score between the adversary's intention and the perceived intention by $B$. From an information theoretical perspective, an interesting analogy is to consider how many bits it takes to change someone's perception of the information.

\subsubsection{Fake media prophylactics}

Suppose the information receiver $B$ was ``exposed'' to the fake media. From a prophylactic perspective, we would like to recommend a set of practices to the receiver $B$ so that they can better recognize the intention of disinformation sender and is prevented from infection. Using the same objective function as in the case of antigens, here, $Y$ represents a set of prophylactic practices, instead of arguments. The objective is to choose a set of practices which maximize the similarity score between the adversary's intention and the perceived intention by $B$.  Discovering, exploring, and testing possible prophylactic practices may boil down to changing this decoder function of $B$.

\subsection{Creating the Theory of Disinformation Consumption}
\label{subsection:create-theory-dis-consumption}


Building a comprehensive and formal theory of disinformation consumption is, in our view, a formidable task, since it requires contributions from multiple disciplines and very different types of studies and experiments. 

One of the key components is to gain a better understanding of how human beings perceive truthfulness and respond to false and deceiving contents. In the specific context of fake news, recent studies have been focused on understanding why false news stories gain popularity on social media. Political partisanship explains part of the attention, in which people uncritically accept arguments that support their political ideology~\cite{thaler2019fake}. Individuals do not neutrally interpret information, tending to favor information that confirms existing beliefs or their world-views~\cite{Lazer1094, Cook17Neutral}. Fact-checking efforts are not an effective solution since it deals with hard evidence and data but it also needs empathy~\cite{empathyFactCheck}. 

Moreover,  some people are still resistant to accept corrections in what they believed even in the presence of evidence of their misconceptions~\cite{Cook17Neutral}. This is not only because that misleading information can be aligned to the belief system (i.e., partisanship, ideological view, religious background, and the like) but also if the context is a frightening scenario affecting people's emotion states (i.e., virus pandemic). 

Recently, some encouraging studies have shown that analytical thinking can decrease self-reported likelihood of ``liking'' or sharing fake news, for instance by asking people to judge the accuracy of a given content before sharing \cite{Pennycook2018PriorEI}. Our approach goes in the same direction to elicit a person's reasoning (i.e., engaging less his/her emotional side) using explanations or arguments which challenge a person's perception of truth.

\subsubsection{Engaging Analytical Thinking} As a way of drawing people's attention and reasoning toward identifying and acknowledging the falsity of a fake media, we will investigate the strategies of eliciting people's analytical thinking. The major challenge of eliciting analytical thinking rest on two aspects of the nature of fake media: (1) the construction of the utterance based on a recognizable character (e.g., your boss, a well-known and well-respected public figure, and the like) (2) its multimodal format. Our hypothesis is that people will naturally resort to Karhneman's \cite{kahneman2011thinking} ``System 1''---pattern recognition---in order to inspect and make sense of fake media, as opposed to evoking ``System 2''---their analytical thinking. Our goal will thus be to create mechanisms that will help elicit ``System 2.'' 

To this end, our approach will, first, explore aspects not addressed in previous work: we will consider not only text but other types of media (audio, video, image)---thus investigating how information is displayed on current social media platforms. In the context of the proposed structure of the theory of disinformation communication \& consumption, we want to explore how the perceived intention of the sender affects perception of falsity, and how arguments about why someone would be sending a piece of fake media could affect someone's willingness to act upon it.

We believe that it is important to focus on understanding what the psychological and cognitive thresholds are to avoid harm, and how they can be triggered and enhanced by different types of arguments. In other words, our research will look into believing as a scaffold to action, by considering the perception of truth as a component of decision theory. By doing so, we may be able to avoid some of the most controversial, and difficult, aspects of understanding and modeling how people's perceptions of reality affect their beliefs and/or acceptances of truth or falsity. We may also be able to avoid some of the perils of persuasion theory~\cite{o2008persuasion}, communication~\cite{stiff2016persuasive}, and technology~\cite{fogg1999persuasive}.

\subsubsection{Working with Replicable Experiments} We propose to start our investigations by designing basic, highly repeatable human studies which put participants in a context where believing in a piece of information and/or acting upon it is associated to some reward or harm (for instance, less economic reward). After firmly establishing the validity of the experimental setup, and the reliability of replicating it in large scale, possibly in a \emph{Mechanic Turk} fashion, we will be in a position to explore how different variables influence the participants' beliefs and behaviours.

In addition, in having a established experimental framework allows us to more easily test the different prevention and cure methods and strategies our research will produce as well as will enable us to compare different techniques proposed by other researchers. We may also use the experimental setup in a large scale so as to be able to produce datasets of fake media samples and related arguments, which may be used in the development of antigen and prophylactic technologies.


\subsection{Developing Automatic Fake Media Detection Methods}
\label{subsection:virus-detection}

Existing deepfake detection methods are mostly focused on leveraging flawed artifacts generated by deepfake techniques. The deepfake techniques are evolving at a fast pace and we need to develop robust detection methods that look beyond flawed artifacts. 
We describe several areas we would like to further explore with the goal to develop \textit{universal} deepfake detection methods that are built on top of fundamental common characteristics of deepfakes, and can be generalized across a broad spectrum of deepfake generators. 


\subsubsection{Spatio-Temporal Tracking of Nuances}
Deepfake media generators alter original media content by replacing or morphing pixels seamlessly to create realistic content. However, when this transformed content is carefully examined manually or via an algorithm, certain nuances (e.g., color pixel or content mismatched) can be detected. The detected nuances in this case are due to spatial changes at the pixel level in certain regions of interest. In the proposed research, we would like to not only look at the spatial subtle differences in color, texture, meaning, etc., but also detect and track the change of these nuances over time. 

One approach to do detection at the spatio-temporal level is to detect unique and similar landmarks across a set of frames. The Scale-Invariant Feature Transform (SIFT) algorithm is one method that offers the ability to find and track certain features or landmarks across an array of images and video frames. The core idea is that unique landmarks found in an initial set of frames, should, theoretically, appear in subsequent frames. For example, if the eye pupil, nose tip, and freckle of a person's face are detected in the first frame, these features should also be detectable in consecutive frames if (1) it is the same person, and (2) if pixels have not been altered and the noise is constant. However, if the variance of pixels of such features varies greatly due to alterations, these features should be easily detectable and tracked across consecutive frames.


This method offers the advantages of being able to track unique features over time, and that no data from different distributions are needed to train a model (e.g., an unsupervised approach). The first set of frames serve as reference frames, and each subsequent detected key landmark across frames form a lattice graph like structure, where each vertex represents a key landmark, and an edge represents the connectivity between consecutive landmarks, frame by frame. A formulation like this could help determine which frames are fake or which regions of a particular frame have been manipulated since the connectivity between landmarks across fake frames tends to be more sporadic. Furthermore, this technique could be applied to create a dataset of interleaving fake and non-fake frames in order to train spatio-temporal network architectures to determine which individual frames or which regions within frames are real or fake.

The SIFT method and others, such as Optical Flow, could be employed to detect inter-frame dissimilarities or discrepancies in frames. Certain properties from existing techniques could be used to create a model that evolves over time as more advanced techniques continue to generate very realistic media content. Detecting and tracking the change of nuances over time is one of the core ideas we would like to further explore, as this could be easily applied or extended to other media content such as audio clips (via spectrograms).

\subsubsection{Biological signals}

Images capture visible wavelengths of light that describe a particular scene with $N$ number of objects. A human's eye cone and rod cells allow for light perception, color differentiation, and perception of depth from images. However, we might not be aware but there exist biological signals hidden within these encoded representations that cannot be seen by the human eye or captured by deep learning algorithms, yet they encode valuable information. These biological signals, also known as \emph{biosignals}, record biological, non-invasive spatio-temporal events related to a chemical, electrical, or mechanical event or activity such as temperature, heart rate, and muscle contraction. We would like to further explore and understand the use of biological signals and how these hidden encoded representations, which are hard to notice and detect in media content, can help detect fake media content, especially since GANs and other deep learning techniques might have a hard time generating or replicating these biosignals.

One of these biosignals that has been explored recently is the blood volume change over time from portrait videos. The work by Ciftci et al., \textit{FakeCatcher} \cite{ciftci2019fakecatcher}, employs the photoplethysmography (PPG) technique to detect optical absorption variations in the micro-vascular bed of tissue during a cardiac cycle \cite{Allen_2007, de2013robust}. The idea is that this biosignal and its signal characteristics can be used as implicit features of authenticity since biosignals are hard to replicate and, in fake media content, they are not temporally and spatially preserved. The use of the PPG technique has been previously explored \cite{wu2012eulerian} to reveal temporal variations in videos, where their method apply a spatial decomposition of the signal with some temporal filtering in order to amplify the signal and be able to visualize the flow of blood from a standard video sequence. 

The use of biosignals can greatly help in detecting fake media content as they are difficult to see with the naked eye and hard to replicate. In a sense, they offer the notion of a "fingerprint" that is unique to a video. The fingerprint can be the cardiac cycle, micro muscle contractions, etc. We would like to further explore this research area in order to create fundamental universal biosignals that could be potentially used as fingerprints, and use them to develop new models that are robust to adversarial attacks or more realistic fake content that might be hard to detect.


\subsection{Developing Fake Media Antigen Creation Systems}
\label{subsection:research-antigen-gen}

In immunology, an \emph{antigen} is a substance which enters a human body and causes the body to make antibodies as a response to fight off disease. We would like to develop methods which can generate antigens for humans to challenge fake media, triggering an immune response. This includes automatically generating arguments for contradicting a piece of fake media or generating explanations of detection mechanisms for fake images, audio, and videos. 

These arguments and explanations should be comprehensible to non-technical users, such as journalists, fact-checkers, policymakers, and ultimately to the general public. To achieve our objective, we will use insights from human psychology to determine how to best present information to a user and we will use machine learning to automate the process of creating arguments or evidence that is able to cause a user to question their false beliefs.

\subsubsection{Measuring Effectiveness of Arguments}
\label{subsubsection:research-arg-eff}
Current argument generation techniques described in section \ref{section:soa-arg-gen} aim to measure the quality of a given argument in order to select the best argument. This requires curated data sets which associate a high rank to arguments that are deemed to be high quality. The ranking of arguments in the data sets described in section \ref{section:soa-arg-gen} are mostly determined by humans who say if the argument seems like it would be convincing or not, and in some instances the human judges which out of two arguments is more convincing.

In our research, we aim to determine argument ranking by evaluating the effectiveness of the argument to change someone's beliefs. This data will be collected from the game described in section \ref{subsubsection:gamification}, where there is a human player acting as an adversary who creates the fake media (which can be considered an argument presented in text and visual format) and a human player who acting as a detector of fake media. From the human creation and detection of fake media, we will collect data about which arguments the adversary thinks will be convincing when they create their fake news, data about what arguments are actually able to convince the detector, and data about which claims of an argument a detector noted as an indication the argument was false. Claims from the argument will be labeled with meta data, such as noting if the argument uses ethos, pathos, or logos.

The data set we want to generate will provide insights into what aspects of a claim make it more believable, what strategies are employed by adversaries in an attempt to mislead, and what results of fake media detection would convince a user a piece of media is fake. This new data set will be helpful in our task to automatically create convincing arguments in the context of fake media and will provide the argument generation community a new data set to use to train their algorithms. 

\subsubsection{Automatic Generation of Antigen}
Automating the process of creating an antigen requires a framework to automate the collection of media and to automate the detection of misinformation. Both will be used as data to create the antigen which may take the form of an argument or an explanation. To collect media we plan to leverage existing technologies, such as the Watson News Discovery service. To detect fake information we will rely on the techniques described in section~\ref{subsection:virus-detection}. 

To create a model that is able to input media and fake media detection results and produce an antigen, we use the field of automatic argument generation as a foundation. This field also aims to automatically generate content that is able to change an individual's perception of a topic. A differentiation of our research compared to current argument generation techniques is that we will use insights gained from human psychology and communication (section~\ref{section:disinformation-theory}). Both of these fields of study provide information about how to effectively reason with an individual which is essential in creating arguments that are able to change a person's perspective on a topic. Additionally, we will use Grice's Theory of Meaning (section~\ref{section:disinformation-theory}) to provide a new metric based on the intentions of a malicious content creator and based on the perception of the intentions from the reader. The metric will allow us to quantify to what extent an individual is able to understand the motives of a malicious user and to what extent an individual is able to determine the validity of a piece of information. To the best of our knowledge, this metric has not been used in the context of fake media which will give us a unique perspective when training models with this metric. 

Creating arguments in the context of fake media, introduces another unique aspect compared to current argument generation techniques. From our research, it has been shown that fake news is often emotionally triggering and there are generally motives, or intentions, for one to create fake news such as financial gain or political persuasion. Knowing this, we can train a model to look for emotionally triggering content and to look for motives of the content creator. With the recognition of this information, we can generate arguments that point out these "red flags" to a user so that the user's analytical mind is triggered and they are better able to rationalize about the piece of fake media presented to them. Our research will explore which arguments prove to be most effective in reasoning with an individuals false beliefs. Examples of arguments that we will explore include providing counter-factual information, explaining the results from detection mechanism, and conveying the intent of the malicious user. This may look like \emph{``X was claimed, however, more credible sources says that X is false and Y is true for reasons M. Moreover, detection mechanisms show with 95\% certainty that X has been manipulated. The reason someone posted X was because they had the intention to do N."} During testing, we will look at the impact of each of these type of sentences to see which shift the perspective of an individual the most, relying on the metrics previously described. 

As a future research direction, information about the prior beliefs of the user could be incorporated into the model creation in order to generate personalized content. We avoid including this in our initial research thread since collecting information about what someone believes, and modeling those beliefs, is a more complex task that is not necessary for automatically generating arguments. However, research shows that taking these prior beliefs into account may increase the ability of the generated content to shift someones perspective on a topic \cite{durmus_cardie_2018}. 
The automatic antigen generator would take those prior beliefs into account and create content which is more aligned to what the user already believes. 



\subsection{Establishing Fake Media Prophylactics Methods}\label{subsubsection:gamification}

A prophylactic is a medication or a treatment designed and used to prevent virus infection. We would like to develop solutions, processes, and practices to raise the disinformation awareness in individuals and educate them on how to distrust fake media. This is a complex and difficult proposal which requires multi-disciplinary efforts and long-term evaluation. 

One of ideas we are exploring is to create and design fake media games which may confer psychological resistance against online disinformation and provide inoculation. Gamification is the process of integrating game mechanics (e.g., fast feedback, collaboration, community, and trust score) to a system in order to motivate participation, engagement, and loyalty \cite{robson2015all}. The primary goal is to maximize engagement to inspire users to learn and remain engaged throughout the process. As previously stated, we would like to provide ways and arguments to challenge the veracity of the content in order to help users throughout the process of overcoming falseness.

Recently, various game-like approaches and studies have been conducted to better assess the user's effectiveness in identifying fake news and deepfakes, while also collecting some additional user behavioral information. Basol et. al \cite{basol2020good} produced a game called \textit{Good News about Bad News}, where players are engaged in the process of learning how to detect fake news by playing the role of the adversary. By exposing players to the mindset behind the adversary, the study suggested that the gamified approach allowed them to provide a psychological inoculation against various fake news manifestations and its strategies for dissemination from an adversarial perspective.

Clever et. al \cite{clever2020fakeyou}, developed a multi-player app called \textit{FakeYou} to investigate and strengthen resiliency against fake news. The aim is to make users aware of fake news, show the challenges in evaluating the presented information, enable them to experience the generation of misleading information, and get deeper understanding into their behavior. They evaluated their system with a small number of participants to get a deeper understanding of the user's behavior and educational effects a game like this may have. In their evaluation study, some participants mentioned that new game rounds sometimes improved their ability  to find correct headlines or that it raised their awareness towards wrong ones.

Another recent study, conducted by the MIT Media Lab, utilized the video dataset from the \emph{DeepFake Detection Challenge}~\cite{dolhansky2019deepfake} to teach players on how to effectively detect deepfakes by showing them two versions of the videos and having them guess which is the fake video. Each stage focuses on a specific attribute for assessing how to spot deepfakes and guides users in teaching them how to spot a particular video as being fake. In addition to the game component, the interface provides a couple of other mechanisms for collecting  additional telemetry such as location of the frame which supported their reasoning for their choices from the first altered frame, the player's emotion, and seeing how user's behaviors change when the video has been flipped upside down.

In our study, we propose to synthesize elements from these gamified approaches to better understand the mechanisms of the generation and consumption of how fake media and generated deepfakes can manifest together, and correspondingly help to shape our thinking behind the Theory of Disinformation. For instance, we can create a platform where players can either choose the role of a detector or an adversary, similar to the notion behind a GAN architecture but with humans instead of neural networks. In this game, the objective would be for the adversary to fool the detector by generating fake news coupled with deepfakes and other media sources, such as text and other digital sources. On the other hand, the detectors role is to not only decide whether the listed sources (which comprise a mix of true and fake news sources) is real or fake, but also identify which part of the media is real and fake and construct arguments to justify their decisions.

Hence through this platform, we can observe the dynamics between these players to better understand some of the strategies that either teams employ. With additional consensual data collection through mouse movements for spatio-temporal behaviors, and other additional telemetry such as eye gaze tracking, we can better assess some of the strengths and weaknesses behind how humans interact with scenarios where they have to make a decision on the legitimacy of a piece of information. This study is far more extensive than many of the prior art, as deepfakes in the wild often do not appear together with the real sources, instead are folded into a narrative that justifies a harmful intention driven by the adversaries' objective to disinform the opposing players.

Furthermore, we can leverage this platform not only as an ecosystem for awareness towards deepfakes in fake media, but also as a testbed for us to evaluate the efficacy of various tools which can help augment the detector's ability to spot deepfakes within fake media. 
The inclusion of sensory stimuli such as the integration of graphics, sound, and story-line, as well as having adaptive challenges that evoke suspense and engagement, can lead to an educational platform that engage users to think critically and proactively.
Thus, this provides us a method to collect rich amount of information while also being able to have an active sand-box environment to evaluate solutions that we propose in this paper.

\subsection{Addressing Ethical Questions and Issues}

In the research outlined in this proposal, there is clearly no shortage of situations, contexts, and experiments which require ethical considerations and controls. To start, this work will be dealing often with the transformation of sound and imagery from people, in many cases done with the intuit of creating an altered, negative portrait of them. We will also have to address the ways and reasons people believe in pieces of information, which may require both knowing part of their cultural contexts and beliefs, but also, in some experiments, looking into ways to challenge information which may conflict or support them. Both experiment design and data collection will have to be extremely careful to conform to ethical guidelines.

At the same time, controlling the use and spread of fake media also deals with issues of freedom of speech, the free circulation of ideas, and government control of information. In general, government efforts towards regulation in this are have been cautious for not regulating against misinformation. The debate about those regulation proposals or laws has raised questions from infringing speech guarantees to uncertainty concerning the blurred definition of fake news and media. Legislation attempts are diverse from arresting people who propagate false information to compel tech companies to be more transparent about their anti-disinformation and advertising policies \cite{lawsFakeNews}. 

Similarly, tech companies have also been resistant to fact check and to remove false claims in case of political discourse. However, during the current coronavirus pandemic, because of the lives risk of fake ``cures'',  social media companies have taken more strict actions. Most of them are sharing official data published by government agencies and health authorities from all over the world. Facebook has removed videos from politicians announcing cures or not tested treatments (e.g., Brazilian president Bolsonaro) and any post containing ``corona'' terms is invited to share a link to WHO's myth-busting site. 

For example, Google has partnered with fact-checking agencies enabling users to check data using the search interface in some countries.  During the pandemics, Google has been flooded people's search results about the virus with government alerts and remove videos on YouTube promising supposed cures or reducing the promotion of hoaxes and other harmless related content. Finally, Twitter has been removing dozens of tweets containing fake news about the virus. Unfortunately, these actions do not seem enough to mitigate fake news dissemination~\cite{techFakeNews}.

In this context, it is legitimate to consider the implications of the research proposed in terms of inadvertently fostering further non-ethical controls of information. Moreover, considering specifically some of areas of this proposal, we could see the potential danger of creating theory and methods which could be used both to improve the ability to convince and do harm to people with false information and, alternatively, to make people disbelieve correct information.

To address those and other ethics related dilemmas, we propose to explicitly include in the research plan a theme which looks, considers, studies, and criticizes what is being constructed in the light of possible negative ethical implications. For instance, we have to develop ways to address the trade-offs between open access to virus detection methods and the need to deter adversaries generating fake media, and balance them.

We would like to develop a new disinformation theory which could be shielded, by design, from most instances of misuse, as briefly discussed before. We should also aim to a new theory of consumption with powerful mechanisms to avoid the deceit of people. Nevertheless, we understand how difficult is to avoid unintended consequences in scientific work, and therefore will include in our team researchers whose main function, and skills, will be focused in mapping, understanding, and pointing to the team, as well as the scientific community in the area, potentially dangerous or ethically dubious ideas and pathways.

\section{Final Discussion}

This paper alerts to the emergence of a new wave of disinformation, one which will use ``cheap" deepfake technologies to create false content to deceive and harm people in a fine-grained manner. This new breed of deepfakes will deceive people by disguising falsity. Unlike today's fake news problem, just fighting its spread is not going to be enough to avoid negative impacts. We thus propose to look into how to automatically help users, who are exposed to fake media, distrust and challenge the veracity of what they are seeing, hearing, or reading. We need to develop automatic technologies to support prevention and cure from contact with fake media.

We argue for the need of a new theoretical framework to understand how fake media is generated, communicated, and consumed, which we call a new \emph{Science of Disinformation}. Our initial idea is to structure such theories using the Shannon-Weaver Information Theory and Grice's Theory of Meaning, and discussed some possible ways to do so in this paper. Although we acknowledge that this is a considerable scientific endeavour, the recent advances in representation of latent spaces may provide an important foundation to build from. The paper also provides a brief survey of current techniques and methods related to generating and detecting fake media, and on machine argumentation technology which is necessary to create effective ways to automate the prevention and cure process. We finish by outlining a research proposal which covers and address the research questions and challenges described in this paper. In addition, we included another important theme relative to possible ethical issues raised by the proposed new theories and technologies.

Although we believe that many of the fruits of the research proposed here will be mostly applicable to the problems the second wave of disinformation, we also expect that key parts of the technology and its associated theories will contribute to the solution of problems related to the spread of fake media in social media networks. In fact, since the solutions for long-tail fake media issues have to depart from the traditional methods of combating fake media, that is, curbing its spread, we hope the research outlined here will provide fresh perspectives and tools in that arena.

Finally, we firmly believe that this is a difficult and complex problem which will have to be addressed in collaboration with the scientific community, open source developers, fact-checking organizations, and other stakeholders of the society, including governments. Fake media is already a global problem and only with combined efforts from multiple actors will prepare us to properly defend ourselves from the upcoming second wave of disinformation.

Deepfakes pose a significant threat to individuals, businesses, and society at large. It will fundamentally alter and affect how we deal with the question of truth in our everyday affairs, profoundly transforming the ways in which we interact with one another via digital technology. 
We plan to investigate and devise new technologies to fundamentally address the problem, not by means of impairing the spread of disinformation, but protecting individual people, organizations and institutions from being deceived by those attacks. We are at the dawn of this new somewhat dytopian reality, and we ought to act to avoid this coming to pass.

\section*{Acknowledgement}

We would like to acknowledge the value and substance of the discussions we had with Rachel Bellamy, Bianca Zadrozny, and Jeanette Blomberg during the writing of this paper.

\bibliographystyle{ACM-Reference-Format}

\bibliography{reference}


\end{document}